\documentclass[a4paper,12pt]{article}
\usepackage[cp1251]{inputenc}
\usepackage[english]{babel}
\usepackage[T2A]{fontenc}
\usepackage{amsmath}
\usepackage{amssymb}
\usepackage{graphicx}

\def\Ha{\hbox{H$_\alpha$\,}}
\def\degr{\hbox{$^\circ$\,}}

\def\Hb{$\mbox{H}_\beta$\,}

\hoffset=+5mm \voffset=+3mm \Roman{table}

\newcounter{tapp}

\def\Ha{\hbox{H$_\alpha$\,}}
\def\degr{\hbox{$^\circ$\,}}

\begin{document}

\begin{table}[t]
\begin{tabular}{|p{17cm}|}
\hline
\it{Astronomy Letters, 2007, vol. 33, No.8, pp.520-530} \\
Translated from Pis'ma v Astronomicheskii Zhurnal, 2007, vol. 33,
No.8, pp.585-597 \\
\hline
\end{tabular}
\end{table}

\begin{center}
{\Large \bf{2D Spectroscopy of Candidate Polar-Ring Galaxies: \\
I. The Pair of Galaxies UGC 5600/09}}

\vspace{0.5cm}
     {\bf L.V. Shalyapina\footnote{lshal@astro.spbu.ru},
     O.A.Merkulova\footnote{olga\_merkulova@list.ru},
     V.A.Yakovleva, and E. V. Volkov}

     { Sobolev Astronomical Institute, St. Petersburg State University, \\
     Universitetskii pr. 2, Petrodvorets, 198904 Russia}

\vspace{0.5cm}

\abstract{
Observations of the pair of galaxies VV 330 with the SCORPIO
multimode instrument on the 6-m Special Astrophysical Observatory telescope
are presented. Large-scale velocity fields of the ionized gas in \Ha and
brightness distributions in continuum and \Ha have been constructed
for both galaxies with the help of a scanning Fabry Perot interferometer.
Long-slit spectroscopy is used to study the stellar kinematics.
Analysis of the data obtained has revealed a complex structure
in each of the pair components. Three kinematic subsystems have been
identified in UGC 5600: a stellar disk, an  inner gas ring  turned with respect
to the disk through $\sim$80\degr, and an outer gas disk. The stellar and
outer gas disks are noncoplanar. Possible scenarios for the formation
of the observed multicomponent kinematic galactic structure are considered,
including the case where the large-scale velocity field of the gas is represented
by the kinematic model of a disk with a warp. The velocity field in the second
galaxy of the pair, UGC 5609, is more regular. A joint analysis of the data
on the photometric structure and the velocity field has shown that this is
probably a late-type spiral galaxy whose shape is distorted by the gravitational
interaction, possibly, with UGC 5600.

}
\end{center}

\section{ Introduction }

Polar-ring galaxies (PRGs) constitute a rare class of dynamically peculiar systems
in which a ring or a disk of gas, dust, and stars rotates around the main body
approximately in the polar plane (Whitmore et al. 1990). The formation of a polar ring
is widely believed to be related to the interaction between galaxies or even
to their mergers. Attempts to model such processes have been made in several works
(see Reshetnikov and Sotnikova 1997; Bekki 1998; Bournaud and Combes 2003).
Despite the progress in interpreting the morphology of the observed objects,
PRGs, achieved in the above works, it should be recognized that there is most likely
no single universal mechanism that would explain the polar-ring formation
in each specific case. Therefore, it is appropriate to perform numerical simulations
to reproduce the observed morphology and velocity field for each of the candidate PRGs.

Much observational data on PRGs have been accumulated to date, but many
questions related to the polar-ring formation, stability, and age still remain debatable.
New, more complete and accurate properties of the stellar population and
the interstellar medium, and the star formation processes is needed to solve
these and several other problems. Optical data on the motion of the gas and stars
are usually obtained with long-slit spectrographs. The results of such observations are
 not always interpreted unambiguously, particularly in the case of complex
 multicomponent systems.

 Analysis of the velocity fields gives much more information.
 The velocity fields are constructed mainly from observations of neutral hydrogen
 or molecular gas (mostly in CO lines) in the radio frequency range and from
 2D or 3D spectroscopy in the optical range. 2D spectroscopy makes it possible
 to study the motions of the gas and stellar subsystems in galaxies in greatest
 detail and to obtain data on a qualitatively new level. However, such data
 for PRGs and similar objects are still very scarce. For example, the velocity fields
 for the best known PRGs, such as NGC 2685 (Shane 1980) and NGC 4650A
 (Arnaboldi et al. 1997), and some other objects were constructed from 21-cm
 observations, but the spatial resolution of these data is low. Until recently,
 the velocity fields in the optical range were obtained only for the central regions
 of two galaxies, NGC 2685 and IC 1689 (Sil'chenko 1998).

 The catalog of
 PRGs byWhitmore et al. (1990) includes 157 galaxies and the existence
 of two roughly orthogonal kinematic systems (classical PRGs) has been
 confirmed observationally only for 15--20 of them. Obviously, the discovery
 of each new object and a detailed study of galaxies from Whitmore's catalog,
 particularly by means of 2D spectroscopy, are of great interest.

 In 2000,
 a program to investigate PRGs and related objects from the catalog of
 Whitmore et al. (1990) by means of 2D spectroscopy using a multipupil
 fiber spectrograph (MPFS) and a scanning Fabry Perot interferometer (FPI)
 of the 6-m Special Astrophysical Observatory (SAO) telescope was initiated
 at the Astronomical Institute of the St. Petersburg State University. One of the
 first interesting results of this program was the detection of a superwind
 from the galaxy NGC 6286 (Shalyapina et al. 2004a). At the same time,
 two almost perpendicular gas systems were found in the galaxy NGC 7468
  (Shalyapina et al. 2004b) and the existence of an inner polar ring was assumed
  on the basis of a joint analysis of 2D spectroscopy and photometry. MPFS data
  on the stellar and gas kinematics in the central regions of candidates PRGs
  confirmed the existence of two almost orthogonal kinematic systems in the central
  region of the galaxy UGC 5600 (Shalyapina et al. 2002). Two stellar kinematic
  systems were discovered in the galaxy UGC 4892, which suggested
  the presence of a satellite (Hagen-Thorn et al. 2003). The velocity field in \Ha for
  the galaxy NGC 2685 was constructed from FPI observations
  (Hagen-Thorn et al. 2005).

\begin{figure}[ht]
    \vspace*{-0.0cm}
    \hspace*{-0.0cm}
    \vbox{ \includegraphics{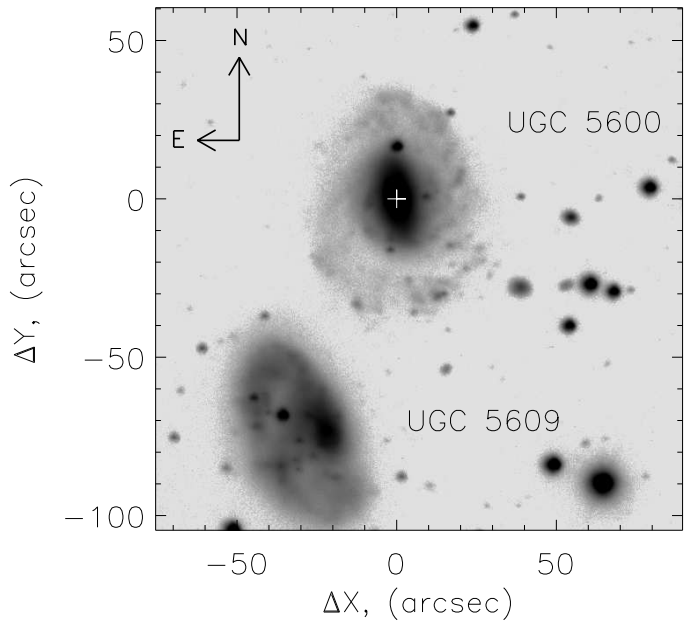}
               } \par
\vspace*{7.5cm} \hspace*{-0.0cm} \caption{\small $R$-band image of
the pair of galaxies VV 330 obtained with the 6-m SAO telescope.}
\label{f:f_1}
\end{figure}

  In this paper, we present FPI observational data
  and new results of long-slit spectroscopy for the isolated pair of galaxies VV 330
Vorontsov--Velyaminov 1959, 1977). An $R$ band image of this pair is shown in
Fig. \ref{f:f_1}.
The scale is 0.2 kpc per 1$''$ if the distance to one of the components, UGC 5600,
is assumed to be 44.6 Mpc at a galactocentric velocity V$_{gal}$ = 2897 km/s and
$H_0$ = 65 km/s/Mpc. Both galaxies have close line-of-sight velocities and
the separation between them in the plane of the sky is 1$'$.4. One of the galaxies,
UGC 5609, with a fairly complex structure, has not been studied in detail previously.
The other galaxy, UGC 5600, is known as the most likely candidate for PRGs
(Whitmore et al. 1990). Our previous study of the stellar and gas kinematics
in the galaxy based on MPFS observations (16$''$ $\times$ 15$''$ field) in the green
and red spectral ranges and on long-slit spectroscopy in the red spectral range
consisted largely in analyzing the velocity field of the central region in this object
(Shalyapina et al. 2002). In this paper, we established that the galaxy
has a complex kinematic structure:
the stars rotate around the galaxy's minor axis, while the gas, except for the
central 3$''$ region, rotates around its major axis. This led us to the conclusion
about the existence of an inner polar ring. However, the picture of the gas motion
in the outer regions remained unclear.

That is why the main objective at this stage
of our study was to obtain information about the velocity field on the scales
of the entire galaxy, including its outermost parts, to compare the new observational
data with the results for the central region, and to bring all data to a unified picture.
In connection with this formulation of the problem, the main emphasis was shifted
to an investigation with FPI. The results of studying the stellar kinematics at a
distance up to 18$''$ from the center also turned out to be very useful at the final
stage. Our more thorough and comprehensive investigation showed that
UGC 5600 is not just the main galaxy with an inner polar ring, but is a more
complex dynamical system and what was previously taken as the inner polar ring
is one of the components of the overall complex structure. This issue is discussed
in detail in Section ``Kinematics of the Gas and Stars''. Until then, we will use the term
``inner ring'' following Shalyapina et al. (2002).

\section{Observations and data reduction}

The spectroscopic observations of the galaxies UGC 5600/09 were performed
at the prime focus of the 6-m SAO telescope with the SCORPIO focal reducer
(Afanasiev and Moiseev 2005). A log of observations for this pair of galaxies is given
in the Table \ref{t:obs}, which also provides information about the $R$-band image. Parameters
of the focal reducer in interferometric observations are given in Moiseev (2002).
Premonochromatization was made using a narrowband filter with $FWHM$ = 20\AA.
The FPI spectral resolution was about 3\AA ~($\sim$130 km/s). Since the
 (CCD TK1024) detector readout was performed in 2 $\times$ 2-pixel binning mode,
 a 512 $\times$ 512-pixel image at a scale of 0$''$.56 per pixel was obtained in each
 spectral channel.

 \begin{table}[ht]
\begin{center}
\caption{Log of observations}
\label{t:obs}
\vskip0.3cm
\begin{tabular}{c|c|c|c|c|c|c}
\hline
\hline
Object & Instrument, & Exposure, & Field & Seeing  & Spectral & Р.А. \\
                & date        &  sec           &          &               &  region, \AA & ~ \\
\hline
UGC5600/09  & IFP 05.03.2002     & 32$\times$180 & 5$'$$\times$5$'$ & 1.8$''$ & \Ha &  \\
UGC5600/09  & SCORPIO 05.04.2002     & 540 & 6$'$$\times$6$'$ & 1.6$''$ & R &  \\
UGC5600  & Slit 23.02.2006      & 11$\times$1200 & 1$''$$\times$6$'$ & 3.0$''$ & 3900--5700 & 0\degr \\
UGC5609  & Slit 21.02.2006      & 4$\times$900 & 1$''$$\times$6$'$ & 1.5$''$ & 3900--5700 & 57\degr \\
\hline
\hline
\end{tabular}
\end{center}
\end{table}

 The interferometric observations were reduced with software developed at SAO
 (Moiseev 2002). After the primary procedures (the subtraction of night-sky lines and
 the reduction to the wavelength scale), the observational data constitute ``data cubes''
 in which each point in a \hbox{512 $\times$ 512-pixel} field contain a 32-channel spectrum.
 Optimal data filtering, a Gaussian smoothing in spectral coordinate with $FWHM$
 equal to 1.5 channels and a 2D Gaussian smoothing in spatial coordinates with
$FWHM$ = 2 pixels, was performed using the ADHOC software
package.\footnote{The ADHOC software package was developed by J. Boulesteix
(Marseilles Observatory) and is freely available on the Internet.}
Gaussian fitting of the \Ha emission line profiles was used to construct the velocity
fields and monochromatic images. The measurement errors of the line-of-sight
velocities for lines with symmetric profiles were $\sim$10 km/s. We also constructed
an image in continuum near \Ha.

We used the method of inclined rings (Begeman 1989; Moiseev and Mustsevoi 2000)
to analyze the velocity field. This method allows us to determine the positions
of the dynamical center and the dynamical axis, to refine the inclination of the galaxy
to the plane of the sky, and to construct the rotation curve. Analysis of the
dependence of the position angle of the dynamical axis and the inclination on radius
provides information about the gas motion in the galaxy. The MPFS data obtained
previously by Shalyapina et al. (2002) were used to convert the \Ha fluxes to the
absolute energy scale.

The slit-spectrograph observations were carried out with the same SCORPIO
instrument in February 2006. The slit length and width are about 6$'$ and 1$''$,
respectively, the scale along the slit is 0$''$.36 per pixel, and the spectral resolution is
5--6\AA. The spectral range contained the \Hb, [OIII] 4959, 5007 \AA ~emission lines and
absorption lines of the old stellar population: MgI 5175\AA, FeI + Ca 5270\AA, and
others. The detector was a 2048 $\times$ 2048-pixel EEV 42--40 CCD array. The data
obtained were reduced using standard procedures of the ESO-MIDAS package.
After the primary reduction, we performed a smoothing along the slit with a
rectangular window 3 pixel in height to increase the signal-to-noise ratio.
The line-of- sight velocities of the gas were measured from the positions of the
centers of the Gaussians fitted to the emission lines. The accuracy of these
measurements was estimated from the night-sky [OI] 5577\AA ~line to
be \hbox{10--15 km/s.}
A cross-correlation method (Tonry and Davis 1979) was used to determine the
line-of-sight velocity and the velocity dispersion from absorption lines. As the
line-of-sight velocity standards, we took the spectra of F8--K3 stars on the same
nights as the galaxy's spectra.

\section{The morphology of UGC 5600/09}

According to the Lyon--Meudon Extragalactic Database (LEDA), the morphological
types of the galaxies UGC 5609 and UGC 5600 are Sbc and S0?, respectively.
Based on a detailed photometric study of UGC 5600 in the $B$, $V$, $R_c$ color
bands, Karataeva et al. (2001) suggested that the galaxy could be most likely
classified as a late-type (Scd) spiral. The presence of such different estimates of the
galaxy type ia most probably explained by the fact that UGC 5600 is a
multicomponent object in which, as will be demonstrated here, several kinematic
subsystems are clearly identified.

\begin{figure}[ht]
    \vspace*{-0.0cm}
    \hspace*{-0.0cm}
    \vbox{ \includegraphics{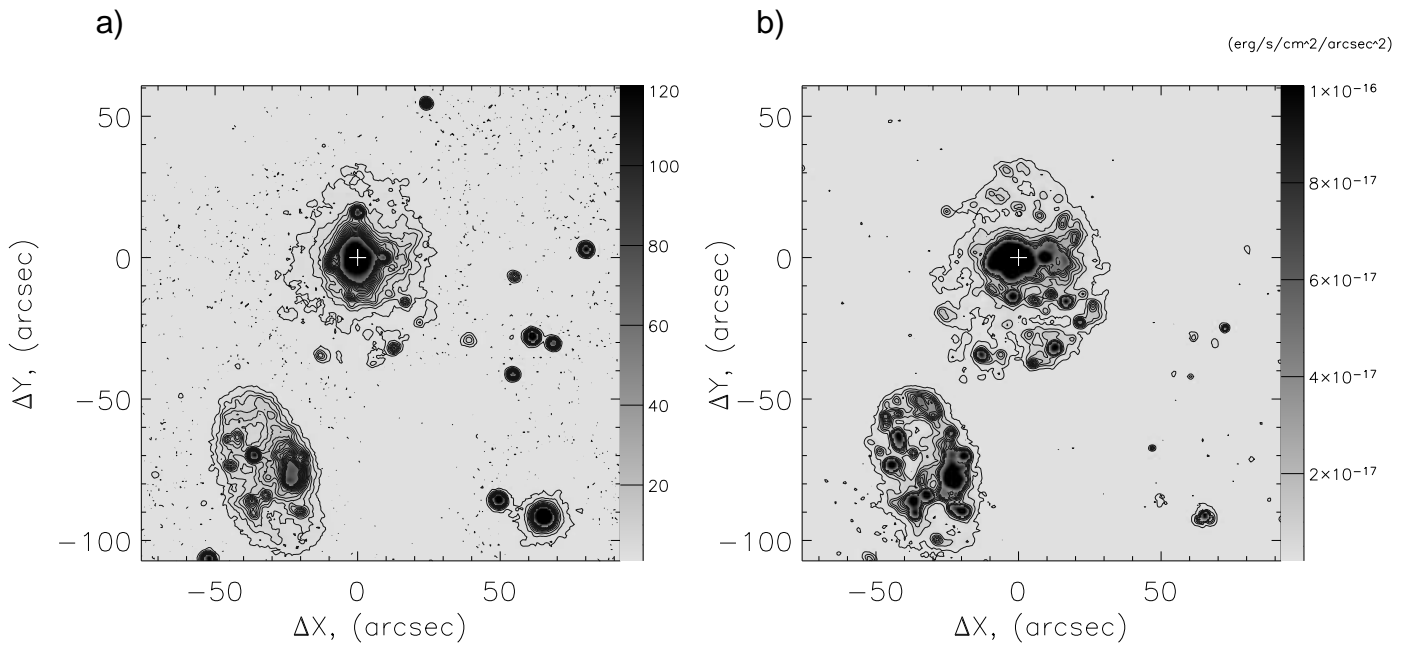}
               } \par
\vspace*{8.cm} \hspace*{-0.0cm} \caption{\small Brightness
distributions for VV 330 obtained with FPI: (a) in a narrow
continuum near \Ha and (b) in \Ha. The gray scale corresponds to
the brightness in arbitrary units for Fig. 2a and in
erg/s/cm$^2$/arcsec$^2$ for Fig. 2b.} \label{f:f_2}
\end{figure}

Figure \ref{f:f_2} shows the continuum (near \Ha) and \Ha brightness distributions for the
binary system VV 330. Note that both galaxies are rich in gas and the \Ha emission is
traceable in them up to the outermost continuum isophotes. Along with diffuse \Ha
emission, both galaxies exhibit numerous bright knotes that are probably star-forming
regions. The central region of UGC 5600 is distinguished by a particularly intense
emission. Let us consider in more detail the brightness distribution for each galaxy.

{\bf{UGC 5600}}.
Comparison of Fig. \ref{f:f_1} with Figs. \ref{f:f_2}a and \ref{f:f_2}b shows that the
continuum
and \Ha brightness distributions are different, particulary in the central region. The
main body of the galaxy, which is an inclined disk whose major axis has a size of
$\sim$30$''$  ($\sim$6 kpc) and a position angle of $\sim$180\degr, is identified in
continuum. In the disk, the isophotes are nearly elliptical in shape with small twists
n the southern side. Protrusions are noticeable on the isophotes in the E--W direction
almost perpendicular to the major axis of the galaxy's main body at a distance of
$\sim$12$''$  from the center. It is this photometric feature that Whitmore et al. (1990)
called the  inner ring.  The main body of the galaxy is surrounded by an extended
outer round envelope with a clumpier structure on the side facing the companion.
Such features in the $B, V, R$ brightness distribution were also pointed out by
Karataeva et al. (2001); they also pointed to a decrease in flattening for isophotes
fainter than 23.5$^m$/arcsec$^2$ in the $B$ band and to a turn of the major axis of the
ellipses fitted to these isophotes through $\sim$10\degr to the west. These authors
also considered the photometric profiles of the galaxy's main body whose shape
does not show the presence of a noticeable bulge, while the brightness distribution
along the major axis in the range from 3$''$ to 18$''$ is well represented by an
exponential disk with the following parameters: $\mu_{0,B}$ = 19.7$^m$/arcsec$^2$
and $h_B$ = 5$''$ (1 kpc).

The \Ha image clearly shows an elongation of the isophotes in the E--W direction
in the central part of the galaxy. The existence of this preferential direction seems
to be not accidental and to be related to the previously noted feature of the
continuum isophotes (the ``inner ring''). The details of the structure of the central
region are better represented in the MPFS data and were described by Shalyapina
et al. (2002). Previously, we already noted that the luminous gas is traceable roughly
to the same boundaries as the outer continuum isophote, but the structure of the
outer gas disk is more complex and contains numerous knots of different brightness
and sizes. Particularly bright knots with sizes up to 5$''$ (1 kpc) are located in the
southern part of the galaxy, with some of them forming chains. These are most
likely stellar complexes or giant HII regions. The \Ha flux in the central region of the
galaxy (r $\le$ 15$''$ (\hbox{3 kpc})) is \hbox{2.5 $\times$ 10$^{-13}$ erg/s/cm$^2$} and the total flux
within the 10$^{-18}$erg/s/cm$^2$/arcsec$^2$ isophote is
\hbox{3.8 $\times$ 10$^{-13}$ erg/s/cm$^2$.} The lower limit for the star formation rate
(without absorption) obtained using a relation from Kennicutt (1998) for the central
region of UGC 5600 is \hbox{0.4 $M_{\odot}$/yr} and SFR = 0.7 $M_{\odot}$/yr for the entire
galaxy.

Whitmore et al. (1990) suggested that the faint outer envelope in UGC 5600 is a ring.
A more detailed study of this outer structure (Karataeva et al. 2001) showed that this
may be not a ring but two spiral arms wounding counterclockwise around the galaxy's
main body. To test this suggestion, we used a  spirality criterion  based on the
expansion of the surface brightness distribution into a Fourier series of the azimuth
angle in the galactic plane followed by analysis of the shape of the lines of the
maxima of individual Fourier harmonics (Moiseev et al. 2004). For our analysis,
we used a deeper (approximately by 0.5$^m$/arcsec$^2$) $R$-band image that
we obtained with SCORPIO (Fig. \ref{f:f_1}) than that in Karataeva et al. (2001). The
amplitude of the second harmonic turned out to be largest; a two-armed structure is
satisfactorily fitted to the observed brightness distribution at the following parameters:
the position angle (PA) of the outer structure 160\degr and the inclination ($i$) of the disk
plane to the plane of the sky 25\degr. Similar values of PA and $i$ for the outer
isophotes in the $B$ band were obtained by Karataeva et al. (2001).

{\bf{UGC 5609}}. The continuum and $R$-band images of the galaxy (Fig. \ref{f:f_2}a
and
Fig. \ref{f:f_1}) show a faint, nearly elliptical disk with a major axis a = 62$''$  ($\sim$12.4 kpc),
a position angle PA = 20\degr, and an axial ratio of 0.5. A slight increase in brightness
is observed in the outer part of the disk and we get the impression that a faint
envelope with a more blurred inner edge is present. No rise in brightness is noticeable
near the geometrical center of the isophotes. Only at a distance of $\sim$8$''$ to NE
is a bright compact feature with nearly spherical isophotes distinguished.
Another prominent feature that stands out against an amorphous disk is located
to the west of the center. This is a fairly bright extended region elongated in the N--S
direction. It has no clear boundaries and its approximate size is 10$''$ $\times$ 15$''$
(2--3 kpc). Since the structure of this region is complex, the isophotes in it are highly
irregular in shape.

The infrared image of the galaxy in the $J,H,K$ bands (2MASS)
differs sharply from the previously described continuum image. Instead of the
amorphous disk, only two bright regions coincident with the features noted in
continuum are seen in all three bands. According to our estimates of the 2MASS data
(we used the calibration from Jarrett et al. 2000), the \hbox{$K$ magnitude} of the bright
compact feature is $\sim$12$^m$ and its color index $J-H$ $\approx$ 0$^m$.9.
Its spectrum turned out to be typical of a late-type  M1--M3? star with a radial velocity
of about  --50 km/s. Thus, this is a Galactic foreground star that is projected onto
the disk of UGC 5609 by chance. The surface brightness of the second, more
extended region ranges from 18$^m$/arcsec$^2$ at the center to 20$^m$/arcsec$^2$
on the periphery. The emission in the $J,H,K$ bands is produced mainly by an old
stellar population, which is usually grouped in galaxies symmetrically about the
dynamical center. Therefore, we assume that this extended region is probably the
main body of the galaxy. This conclusion is also confirmed by the analysis of the
kinematics performed below.

The ring envelope, the absence of a bright center, and the bright knot on one side
of the ring all resemble classical collisional ring galaxies, for example, II Zw 28
(Appleton and Marston 1997). However, if we examine in more detail the images
of this galaxy in different color bands (e.g., Fig. \ref{f:f_1} and the blue image from SDSS),
then we can notice that UGC 5609 more likely resembles a spiral galaxy whose
structure is distorted by tidal interaction, especially in the blue band. The bright knot
is the main body of the galaxy from which an arc-shaped spiral arm or tidal tail goes
away to the north. The second arm goes to the east and then turns south at a
distance of about 10$''$ from the geometrical center, forming the SE part of the
luminous envelope.

The \Ha image of the galaxy (Fig. \ref{f:f_2}b) also has a roughly elliptical shape but with
a more blurred outer edge than that in continuum. Numerous knots of various sizes
and brightnesses, which are probably HII regions, are superimposed on the faint
underlying galaxy along the arc-shaped structures mentioned above. A considerable
number of bright knots is also observed in the main body of the galaxy; the \Ha flux
from it is $\sim$ 3 $\times$ 10$^{-14}$ erg/s/cm$^2$. At the same time, the total flux
within the 10$^{-18}$ erg/s/cm$^2$/arcsec$^2$ isophote is
1.3 $\times$ 10$^{-13}$ erg/s/cm$^2$.

\section{Kinematics of the gas and stars}

\begin{figure}[!ht]
    \vspace*{-0.0cm}
    \hspace*{-0.0cm}
    \vbox{ \includegraphics{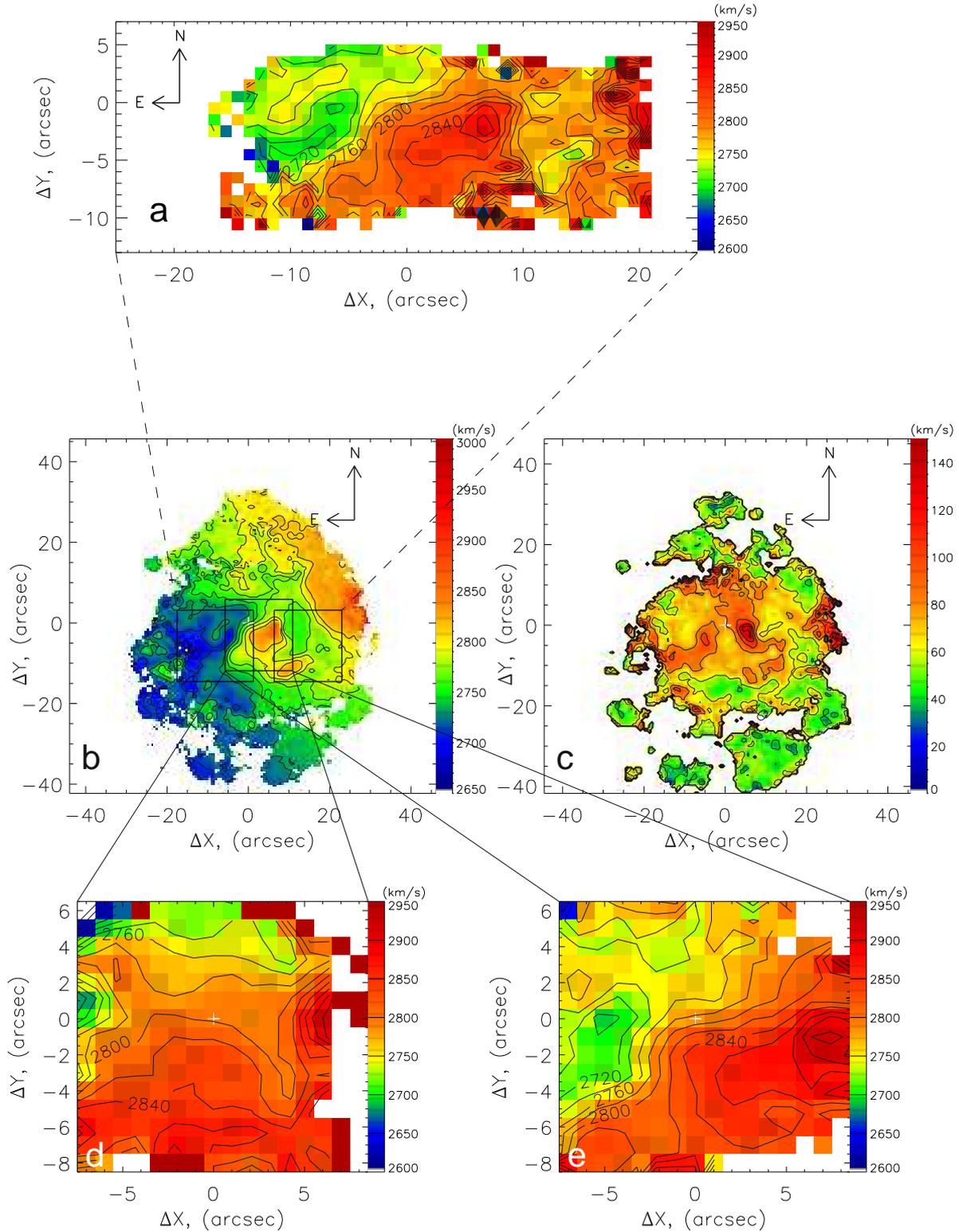}
               } \par
\vspace*{18.5cm} \hspace*{-0.0cm} \caption{\small Line-of-sight
velocity fields for UGC 5600 in \Ha: (a) from three MPFS frames
and (b) FPI data; (c) velocity dispersion distribution. The
velocity fields for the galaxy's central region from MPFS data:
(d) stellar component and (e) ionized gas, \Ha. The step between
the isovels in all figures is 20 km/s.} \label{f:f_3}
\end{figure}

{\bf{UGC 5600}}. Figures \ref{f:f_3}b and \ref{f:f_3}c show the large-scale
line-of-sight velocity field
for the galaxy obtained in \Ha with FPI and the velocity dispersion map. The
distribution of the lines of equal velocity (isovels) shows regular deviations from the
normal pattern of rotation of a flat galactic disk. A kinematic subsystem rotating
around the galaxy's major axis, whose existence was established from MPFS
observations (Shalyapina et al. 2002), is clearly identified in the region of the inner
ring. For the convenience of analyzing the total velocity field of the galaxy, Figs. \ref{f:f_3}d
and \ref{f:f_3}e present the MPFS velocity fields for the inner ring and the central region
obtained from stars and ionized gas. As we see from the figures, the MPFS and FPI
velocity fields in \Ha are in good agreement. In particular, the run of the isovels in the
 nuclear region (r $\le$ 2$''$) in both Fig. \ref{f:f_3}b and Figs. \ref{f:f_3}a and
 \ref{f:f_3}e coincides with the
direction of stellar rotation (Fig. \ref{f:f_3}d).

Previously (Shalyapina et al. 2002), we noted that the system velocities found from
the stars and gas differ. However, our new long-slit observations to be discussed
below did not confirm the differences. Therefore, we once again thoroughly analyzed
the line-of-sight velocities of the stellar component determined from MPFS data and
concluded that they should be increased by 30 km/s. Figure \ref{f:f_3}c shows the corrected
velocity field for the stellar component; the system velocity is V$_{sys}$ = 2770
$\pm$ 10 km/s, which is equal, within the error limits, to the system velocity
determined from emission lines. The velocity field of the stellar component clearly
points to rotation around the galaxy's minor axis; the northern side approaches us,
while the southern side recedes from us. Previously (Shalyapina et al. 2002), we
estimated the position angle (PA$_{dyn}$ = 182\degr). It turned out to coincide with
the position of the photometric axis of the galaxy's main body (Karataeva et al. 2001).
Analysis of the total velocity field for the gas component (Fig. \ref{f:f_3}b) shows that the gas
in the outer parts rotates around an axis inclined with respect to the rotation axis of
the stellar disk (for a more detailed discussion, see below). Moreover, the rotation has
such a direction that the gas in the NW part of the outer envelope recedes from us,
while the gas in its SE part approaches us.

\begin{figure}[!ht]
    \vspace*{-0.0cm}
    \hspace*{-0.0cm}
    \vbox{ \includegraphics{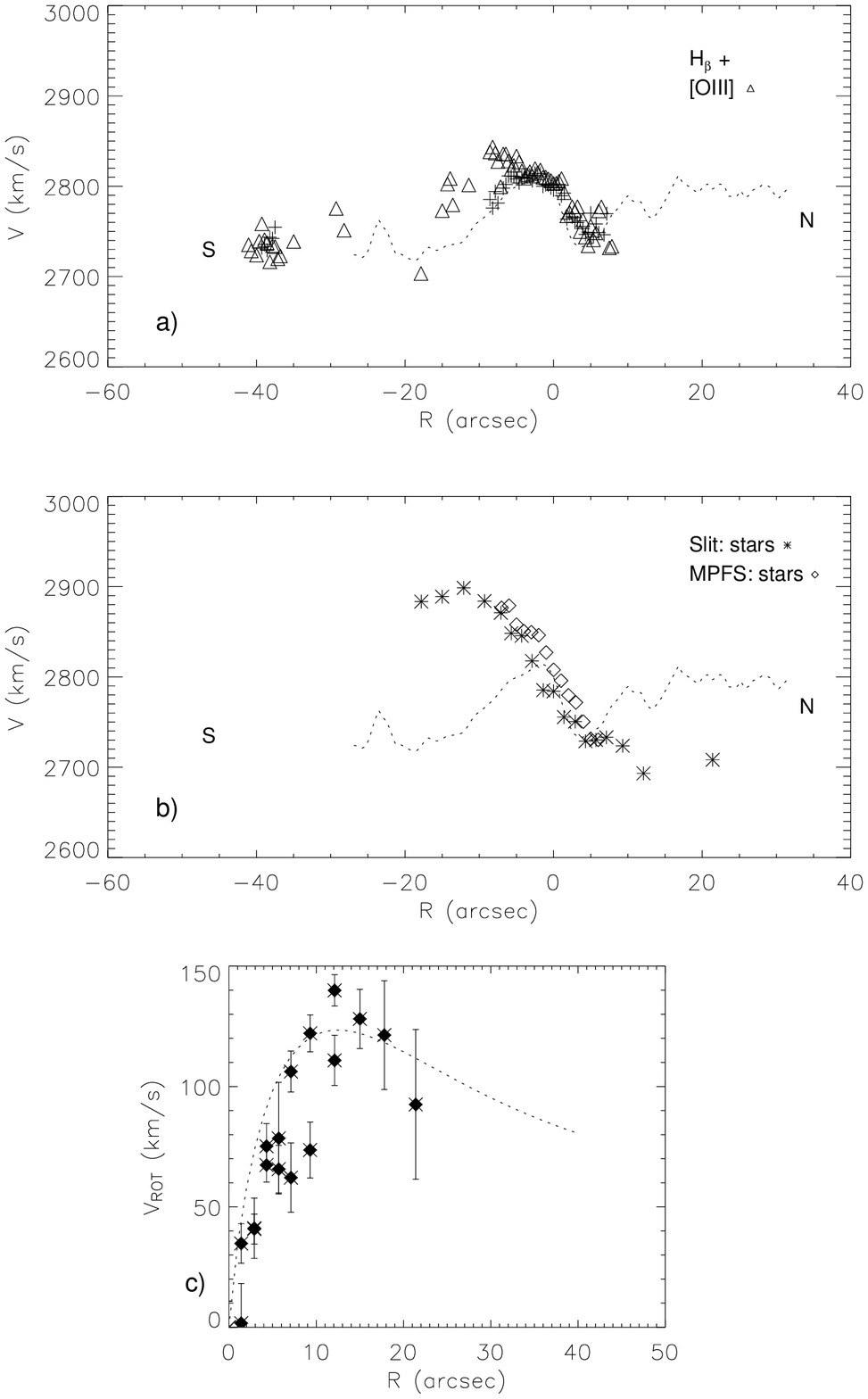}
               } \par
\vspace*{18.5cm} \hspace*{-0.0cm}
\caption{\small Long-slit
line-of-sight velocity curves for UGC 5600 at PA = 0\degr: (a)
ionized gas in \Hb and [OIII] 5007\AA, (b) stellar component, the
dashed lines in both figures indicate the FPI data in \Ha, (c)
rotation curve for the stellar disk (the dashed lines indicate
the model rotation curve of a thin exponential disk).}
\label{f:f_4}
\end{figure}

Since the gas observations on the scales of the entire galaxy revealed a more
complex structure than that followed from the observations of the central region, the
following question arose. How the stars behave at distances from the center larger
than those measured with MPFS? It is with this goal that we performed the long-slit
(PA = 0\degr) observations in the green spectral range. Figure \ref{f:f_4}b presents the
line-of- sight velocity curve for the stars; for comparison, two sections of the velocity
fields for the stellar (MPFS) and gas (FPI) components are also shown here. As we
see from the figure, the long-slit and 2D spectroscopic data are in close agreement,
within the error limits, and the pattern of stellar motion is retained in regions farther
from the center. At the same time, there is a significant contrast in the stellar and gas
kinematics at distances 10$''$ $\le$ r $\le$ 20$''$ from the center. The line-of-sight
velocities of the gas obtained from both permitted \Ha lines (FPI) and forbidden lines
(long slit, Fig. \ref{f:f_4}a) are in good agreement. Figure \ref{f:f_4}c presents the rotation curve for the
stellar component; the theoretical rotation curve of a thin exponential disk with a scale
factor h$_{disk}$ = 1.2 kpc is indicated here by the dashed line. This value is close to
that obtained by Karataeva et al. (2001) from $B$-band photometric data. The mass
of the galaxy within a radius r $\le$ 2.4 kpc is estimated to be
\hbox{$\sim$10$^{10}$ M$_{\odot}$.}

Our analysis of the large-scale line-of-sight velocity field by the method of inclined
rings has shown the following. In the region 3$''$ $\le$ r $\le$ 10$''$, the behavior of
the isovels corresponds to the gas rotation around the major axis of the stellar disk.
In the outer parts at r $\ge$ 20$''$, the dynamical axis of the gas is along the NW--SE
line. Thus, it occupies an intermediate position between the dynamical axes of the
stellar disk and the inner gas ring. This suggests that there are two kinematically
decoupled gas subsystems: an inner ring and an outer disk. The centers of these
subsystems coincide, within the error limits. In the region 10$''$ $\le$ r $\le$ 20$''$, the
structure of the line-of- sight velocity field is very complex and it cannot be reliably
represented by a model of circular motion. Note that the photometric data reveal
a sharp decline in brightness, a change in isophotal flattening, and a turn of the
major axis of the ellipses  fitted to the isophotes in this region.

Therefore, we constructed two separate models of circular motion for the inner
subsystem (3$''$ $\le$ r $\le$ 10$''$) and the outer disk (20$''$ $\le$ r $\le$ 40$''$).
As we have already noted above, the system velocity in both models is the same
and equal to 2770 $\pm$ 5 km/s. For the inner ring, we obtained the following
dynamical parameters: $i_{dyn} \sim$ 60\degr $\pm$ 10\degr and PA$_{dyn} \sim$
260\degr $\pm$ 10\degr. As in our previous paper (Shalyapina et al. 2002), the angle
between the planes of the inner gas ring and the stellar disk was found to be
$\sim$80\degr $\pm$ 10\degr or $\sim$65\degr $\pm$ 10\degr. On the periphery
\hbox{20$''$ $\le$ r $\le$ 40$''$,} the model of gas motion has the following parameters:
PA$_{dyn} \simeq$ 303\degr $\pm$ 7\degr and $i_{dyn}$ = 30\degr $\pm$ 10\degr.
The angle between the planes of the stellar disk and the outer gas disk is estimated
to be 70\degr $\pm$ 10\degr and 40\degr $\pm$ 10\degr.

Below, we make several remarks on the velocity dispersion distribution for clouds
of ionized gas. As we see from Fig. \ref{f:f_3}c, the maximum dispersion is observed in the
nuclear region (more than 100 km/s). As we recede from the center, the dispersion
decreases and is $\sim$50 km/s in the outer parts. In the region of the inner ring,
the dispersion varies between 50 and 100 km/s, but the structure of its distribution
is highly nonuniform and clumpy. It should be noted that there exists a problem
of accurately measuring the velocity dispersion if its value is smaller than the FWHM
of the FPI instrumental profile, $FWHM$ = 130 km/s, which corresponds to
a dispersion $\sigma_{gas} \approx$ 55 km/s. Estimations of the stellar velocity
dispersion along the galaxy's major axis showed that it is lower than the velocity
dispersion of the ionized gas at the center ($\sigma_{\star} \le$ 50 km/s) and
increases to 80 km/s at a distance of 10$''$ from the center.

\begin{figure}[ht]
    \vspace*{-0.0cm}
    \hspace*{-0.0cm}
    \vbox{ \includegraphics{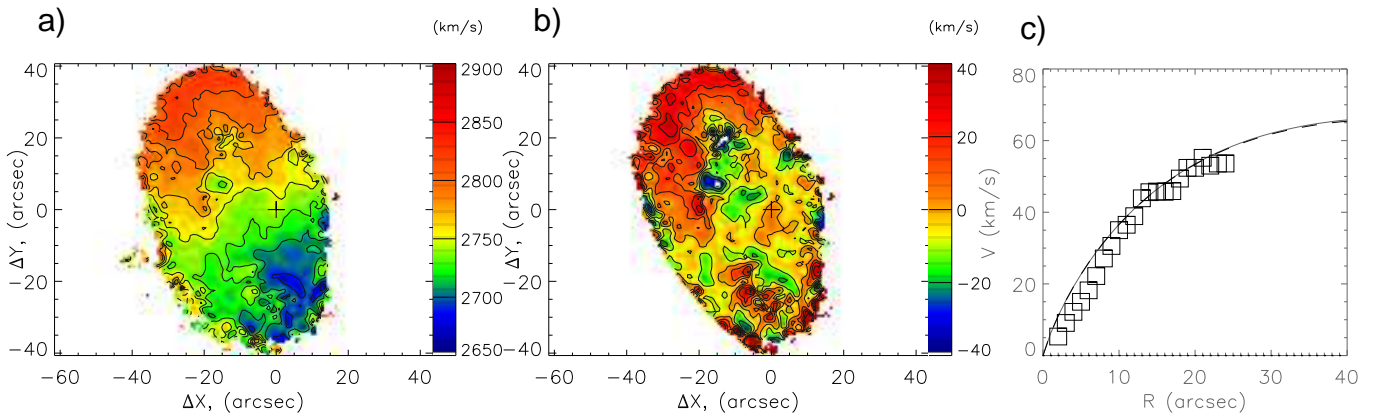}
               } \par
\vspace*{4.8cm} \hspace*{-0.0cm} \caption{\small UGC 5609: (a)
observed velocity field, (b) residual velocity field, (c) model
(solid line) and observed (squares) rotation curves. The step
between the isovels in all figures is 20 km/s.} \label{f:f_5}
\end{figure}

{\bf{UGC 5609}}. Figure \ref{f:f_5}a presents the line-of-sight velocity field constructed
from FPI data in \Ha. At  first glance, it appears regular and similar to the rotation
of the gas disk around the minor axis. A detailed analysis of the field showed that its
structure is more complex; numerous twists that can be caused by various factors
are noticeable on the isovels. First, the structure of UGC 5609 is probably severely
distorted either by tidal interaction or by galaxy collision. Second, the measurement
errors of the line-of-sight velocities depend on the brightness level, while the
brightness distribution in \Ha is highly nonuniform. Therefore, the behavior of the
isovels is related to different errors. Third, to the NE of the geometrical center of the
isophotes at a distance of $\sim$8$''$, the distortions in the velocity distribution are
produced by the absorption component (from a foreground star) in the red \Ha wing.
Since the line-of-sight velocity range for this galaxy is small, these factors will affect
the shape of the isovels.

As we noted above, on the one hand, UGC 5609 is similar to collisional ring galaxies
and, on the other hand, this can be a spiral galaxy whose structure is distorted by
gravitational interaction. Our analysis of the velocity field for this galaxy was also
performed by the method of inclined rings, but we constructed different models: both
with allowance made only for the circular velocity vector component and with the
assumption of nonzero radial velocities (expansion of the outer ring structure). As the
dynamical center, we took either the center of the western knot, which we believe
to be the galaxy's main body, or the geometrical center of the ellipses fitted to the
isophotes. Next, we constructed the model velocity fields and compared them with
the observed field. Note that the residual velocities for most models are low,
\hbox{10--20 km/s.} Since the velocity is almost constant in the central region along the minor axis
of the ellipses, the system velocity is determined unambiguously and its value is
\hbox{2740 $\pm$ 5 km/s;} the inclination does not depend on the model either, $i$ = 55\degr.
If the dynamical center coincides with the geometrical center of the ellipses, then
the maximum positive residual velocities are observed in the region of the main body.
Near the geometrical center (r $\le$ 10$''$), the residual velocities are opposite in
sign and do not exceed 10--15 km/s, but a strong depression is noticeable in the
initial segment (r $\le$ 15$''$) of the rotation curve. The same picture is observed in
the models with a nonzero ring expansion velocity. The residual velocity reaches
+20 km/s on the periphery of the galaxy in the northern and southeastern parts of the
ring. Thus, we find no evidence for the existence of an expanding envelope (ring)
at the level of the measurement errors.

If the dynamical center is displaced by about 10$''$ to the west to the center of the
main body, then the entire velocity field is satisfactorily described by the model of
circular motion. Figure \ref{f:f_5}b presents the residual velocity map. The rotation curve can
be represented by an exponential disk (Fig. \ref{f:f_5}c) with $h$ = 4 kpc; in this case,
a small change in the position of the dynamical axis from 15\degr to 30\degr is
observed and the residual velocities in the region of the  northern  spiral reaches
their maximum value of +30 km/s. If this is assumed to be a tidal structure,
then it can be warped with respect to the plane of the galactic disk. The total mass
of the galaxy is estimated to be 1.6 $\times$ 10$^{10}$ M$_{\odot}$.

\section{Discussion}

Previously, we have already mentioned that the candidate PRGs from the list by
Whitmore et al. (1990) are the objects of our study. However, as we see from the
previous section, no classical polar ring is observed in the galaxy UGC 5600, as,
for example, in the classical PRGs NGC 4650A, NGC 2685, and IC 1689. In this case,
we most likely encounter the phenomenon of what was called  kinematically
decoupled components  in the literature (see Bertola and Corsini 1999). The
phenomenon of counterrotation in several galaxies (Rubin 1994; Galletta 1996),
the circumnuclear polar rings (Sil'chenko et al. 1997; Corsini et al. 2002), and
the classical polar rings (Whitmore et al. 1990) also fall under this name.
Despite the differences in photometric structure and the kinematic peculiarities,
all of them are probably united by the origin: they are the result of (occasionally
multiple) galaxy mergers or mass accretion onto the galaxy from other galaxies.
Let us consider the possible scenarios that unite the observed peculiarities of VV 330.

Since the observed picture for UGC 5600 is very complex, let us summarize the
main observational facts before passing to a discussion.

(1) The main body is a stellar disk with a diameter of $\sim$6 kpc, PA$_{dyn}$ $\sim$
182\degr, and \hbox{$i_{dyn}$ $\sim$ 50\degr} (Shalyapina et al. 2002). The presence of
an ionized gas rotating in the same way as the stars was established in the central
region (r $\le$ 3$''$).

(2) The inner gas ring rotating in a plane close to the polar plane
of the stellar disk has PA$_{dyn}$ $\sim$ 260\degr and $i_{dyn}$ $\sim$ 60\degr.
The pattern of kinematics in the region of the inner gas ring is complicated
by the fact that we possibly see the total emission from the ionized gas that belongs
to the ring and the galactic disk, but we cannot separate them because
the instrumental profile has a large FWHM.

(3) The outer gas disk or ring of low surface brightness is traceable at distances
approximately from 4 to 8 kpc from the galactic center. For this component,
PA$_{dyn}$ $\sim$ 303\degr and \hbox{$i_{dyn}$ $\sim$ 30\degr.} Thus, the outer gas disk
rotates in a plane that is appreciably inclined (either by 70\degr to by 40\degr) to the
plane of the galaxy's main body.

Taking into account the facts listed above,
we considered the following cases.

It can be assumed that UGC 5600 is the projection of two galaxies: one is
a gas-poor polar-ring galaxy seen at a large angle to the plane of the sky and
the other is a gas-rich late spiral. Since the centers of all the observed structures
coincide, within the error limits, and since the system velocities of the stellar and
gas components are close, this case seems unlikely, although it cannot be ruled
out completely.

The next explanation for the observed features of UGC 5600 is
the merger of two galaxies; different scenarios are possible in this case. For example,
the existence of stellar and outer gas disks rotating in different planes can result
from the merger of two (late and earlier-type) galaxies that produced the above
independent kinematic subsystems. The picture is complicated by the presence
of an inner ring. To explain this feature, we can attempt to use a mechanism
suggested by Friedli and Benz (1993) that is related to the presence of a bar
in the galaxy. However, no signatures of an extended bar were found in the
velocity field and the photometric structure for UGC 5600. Recurrent mergers
can serve as an alternative explanation.

In the last case that we will consider,
a warped polar ring is assumed to be present in UGC 5600. One of the
formation mechanisms for such rings is mass accretion from a gas-rich galaxy
onto a neighboring galaxy. A warped polar ring or disk is formed around the
latter as a result of mass transfer (Bournaud and Combes 2003). The second
component of the pair, \hbox{UGC 5609,} may be considered as the donor galaxy.
Let us assume that the entire gas in \hbox{UGC 5600} constitutes a single system and
that the turn of the dynamical axis from the central regions to the periphery is
related to the warp of the gas disk. In this case, the observed features of the
velocity field can be the result of projection effects. To test this hypothesis, we
constructed a model of a warped ring. We used the density distribution in a thin
exponential disk whose rotation curve is shown in Fig. \ref{f:f_4}c. We assumed that
PA$_{dyn}$ and $i_{dyn}$ changed linearly from the inner edge of the ring
to its outer edge. As the first approximation, at the ring edges, we took the angles
obtained by analyzing the velocity field by the method of inclined rings
(see the previous section) for the inner ring and the outer gas component.
We calculated the velocities in the ellipses at given position of the dynamical axis
and inclination. The boundary values of the angles were then varied to achieve
\begin{figure}[ht]
    \vspace*{-0.0cm}
    \hspace*{-0.0cm}
    \vbox{ \includegraphics{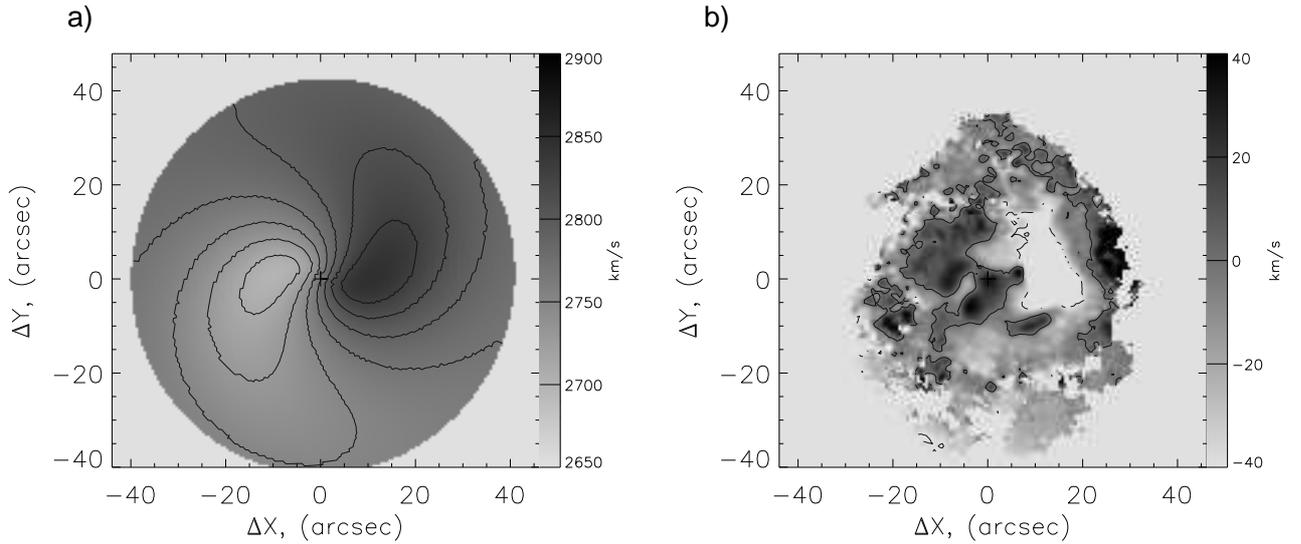}
               } \par
\vspace*{6.cm} \hspace*{-0.0cm} \caption{\small UGC 5600: (a)
rotation model with a disk warp and (b) residual velocities.}
\label{f:f_6}
\end{figure}
the best agreement with the observed velocity field. The model velocity field
shown in Fig. \ref{f:f_6}a was constructed at the following parameters:
PA$_{dyn}$ = 240\degr and $i_{dyn}$ = 45\degr for the inner boundary;
PA$_{dyn}$ = 300\degr and $i_{dyn}$ = 30\degr for the outer boundary.
The angle between the normals to the surface at the inner and outer boundaries is
$\sim$10\degr. As we see from the residual velocity map (Fig. \ref{f:f_6}b), the model
velocity field describes satisfactorily the observed one, except the region
at distances of 10$''$ -- 15$''$ from the center to the west. The feature in the
velocity field is also observed at this location.

At this stage of our study, we do not have sufficient information to choose
between the suggested scenarios.

In conclusion, let us discuss the data obtained for UGC 5609. Based on the
continuum and \Ha images of the galaxy and on their comparison with the image
in the blue band, we made the following assumptions: this is either a collisional
galaxy or a spiral galaxy whose shape was distorted by the gravitational interaction,
possibly, with UGC 5600. Our analysis of the velocity field revealed no expanding
envelope. Based on the entire data set, we concluded that UGC 5609 is most likely
a late-type spiral galaxy with two arcshaped spiral arms; as a result, the illusion
of an envelope is created. One of the arms may be a tidal tail. The high brightness
of this feature in \Ha, the $R$ band, and the blue range and its low brightness
in continuum near \Ha suggest that this arm consists mainly of gas and blue stars,
which cannot be said about the second arm; the latter is much brighter in continuum.
 In addition, our analysis of the velocity field revealed that it is warped with respect
 to the galactic plane.

 As yet no bridges and bars between the galaxies of the pair have been detected,
 but the presence of a tidal tail in UGC 5609 and the distortions of the structure
 on the southern side of UGC 5600 may be indicative of the gravitational interaction
 between the galaxies. We hope to reach ultimate conclusions about the nature
 of VV 330 after numerical simulations of all the possible cases considered above.

 \section{Acknowledgements}

 We are grateful to the Commission on the Subject of Large Telescopes for
 allocating observational time on the 6-m telescope. We also wish to thank
 A.V. Moiseev (SAO RAS) for help with the observations on the 6-m telescope
 and the FPI data reduction, for the provided data analysis codes, and for valuable
 remarks in preparing the text of the paper. This work was supported by the
 Russian Foundation for Basic Research (project no. 05-02-17548) and the
 Ministry of Russian Education (RNP.2.1.1.2852).

\vspace{1cm}

{\large \bf References}
\begin{flushleft}

1. V. L. Afanasiev and A. V. Moiseev, Pis'ma Astron. Zh. 31, 214 (2005)
[Astron. Lett. 31, 194 (2005)].

2. P. N. Appleton and A. P.Marston, Astron. J. 113, 201 (1997).

3. M. Arnaboldi, T. Oosterloo, F. Combes, et al., Astron. Astrophys. 113, 585 (1997).

4. K. G. Begeman, Astron. Astrophys. 223, 47 (1989).

5. K. Bekki, Astrophys. J. 499, 635 (1998).

6. P. Bertola and E. M. Corsini, Galaxy Interactions at Low and High Redshift,
Ed. by J. Barnes and D. B. Sanders (Kluwer, Dordrecht, 1999), p. 149.

7. F. Bournaud and F. Combes, Astron. Astrophys. 401, 817 (2003).

8. E. M. Corsini, A. Pizzella, and F. Bertola, Astron. Astrophys. 382, 488 (2002).

9. V. A.Hagen-Thorn, L. V. Shalyapina, G.M. Karataeva, et al., Astron. Lett.
29, 133 (2003).

10. V. A.Hagen-Thorn, L. V. Shalyapina, G.M. Karataeva, et al.,
Pis'ma Astron. Zh. 82, 1071 (2005) [Astron. Rep. 49, 958 (2005)].

11. G. Galletta, Astron. Soc. Pac. Conf. Ser. 91, 429 (1996).

12. D. Friedli and W. Benz, Astron. Astrophys. 268, 65 (1993).

13. T. H. Jarrett, T. Chester, and R. Cutri, Astron. J. 119, 2498 (2000).

14. G. M. Karataeva, V. A. Yakovleva, V. A. Gagen-Torn, et al.,
Pis'ma Astron. Zh. 27, 94 (2001) [Astron. Lett. 27, 74 (2001)].

15. R. Kennicutt, Ann. Rev. Astron. Astrophys. 36, 189 (1998).

16. A. V. Moiseev, Bull. SAO 54, 74 (2002).

17. A. V. Moiseev and V. V. Mustsevoi , Pis'ma Astron. Zh. 26, 190 (2000)
[Astron. Lett. 26, 565 (2000)].

18. A. V. Moiseev, J. R. Valdes, and V. H. Chavushyan, Astron. Astrophys.
421, 433 (2004).

19. V. Reshetnikov and N. Sotnikova, Astron. Astrophys. 325, 933 (1997).

20. V. Rubin, Astron. J. 108, 45 (1994).

21. L. V. Shalyapina, A. V. Moiseev, and V. A. Yakovleva, Pis'ma Astron. Zh.
28, 505 (2002) [Astron. Lett. 28, 443 (2002)].

22. L.V. Shalyapina, A.V.Moiseev,V.A. Yakovleva, et al., Pis'ma Astron. Zh.
30, 3 (2004a) [Astron. Lett. 30, 1 (2004a)].

23. L. V. Shalyapina, A. V. Moiseev, V. A. Yakovleva, et al., Pis'ma Astron. Zh.
30, 643 (2004b) [Astron. Lett. 30, 583 (2004b)].

24. W.W. Shane, Astron. Astrophys. 82, 314 (1980).

25. O. K. Sil'chenko, Astron. Astrophys. 330, 412 (1998).

26. O. K. Sil'chenko, V. V. Vlasyuk, and A.N. Burenkov, Astron.
Astrophys. 326, 941 (1997).

27. J. Tonry and M. Davis, Astron. Astrophys. 84, 1511 (1979).

28. B. A. Vorontsov-Velyaminov, Atlas and Catalogue of 356 Interacting Galaxies
(Mosk. Gos. Univ, Moscow, 1959) [in Russian].

29. B. A. Vorontsov-Velyaminov, Astron. Astrophys., Suppl. Ser. 28, 1 (1977).

30. B. C. Whitmore, R. A. Lucas, D. B. McElroy, et al., Astron. J. 100, 1489 (1990).

\end{flushleft}

\hspace{10cm}
\it{Translated by V. Astakhov}

\end{document}